\documentclass[runningheads,10pt]{llncs}

\usepackage{graphicx}
\usepackage{amsmath}
\usepackage{amssymb}
\usepackage{amsfonts}
\usepackage[bookmarks]{hyperref}
\usepackage{multirow}
\usepackage{array}
\usepackage{float}
\usepackage{colortbl}
\usepackage{arydshln}
\usepackage{slashbox}
\usepackage{stmaryrd}

\newcommand{\getsr}{\stackrel{{\scriptscriptstyle\$}}{\gets}}

\begin{document}

\title{Privacy-preserving and yet Robust Collaborative Filtering Recommender as a Service}
\author{Qiang Tang}

\institute{Luxembourg Institute of Science and Technology (LIST)\\
4362, Esch sur Alzette, Luxembourg\\
qiang.tang@list.lu
}

\maketitle

\begin{abstract}
Collaborative filtering recommenders provide effective personalization services at the cost of sacrificing the privacy of their end users. Due to the increasing concerns from the society and stricter privacy regulations, it is an urgent research challenge to design privacy-preserving and yet robust recommenders which offer recommendation services to privacy-aware users. Our analysis shows that existing solutions fall short in several aspects, including lacking attention to the precise output to end users and ignoring the correlated robustness issues. In this paper, we provide a general system structure for latent factor based collaborative filtering recommenders by formulating them into \emph{model training} and \emph{prediction computing} stages, and also describe a new security model. Aiming at pragmatic solutions, we first show how to construct privacy-preserving and yet robust \emph{model training} stage based on existing solutions. Then, we propose two cryptographic protocols to realize a privacy-preserving \emph{prediction computing} stage, depending on whether or not an extra proxy is involved. Different from standard Top-k recommendations, we alternatively let the end user retrieve the unrated items whose predictions are above a threshold, as a result of our \emph{privacy by design} strategy. Experimental results show that our new protocols are quite efficient.
\end{abstract}

\section{Introduction}

\begin{figure}
\label{system}
    \centering
    \begin{minipage}{0.4\textwidth}
        \centering
        \includegraphics[scale=0.55]{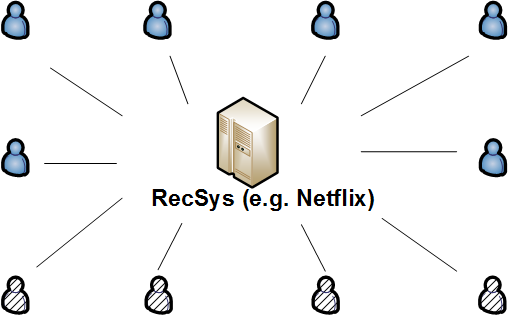} 
    \end{minipage}\hfill
    \begin{minipage}{0.46\textwidth}
Real-world recommenders often build a preference model based on data from a number of sources, such as user's explicit feedback (e.g. rating vectors) and implicit information (e.g. how long a user has stayed on the page of an item). Like in most literature work and for simplicity reasons, we only consider explicit feedback in this paper. The discussions and proposed solutions might also be applied to other types of data.
    \end{minipage}
        \caption{Standard Recommender Structure}
\end{figure}

Today, personalization is widely adopted by a large number of industries, from entertainment to precision medicine. The main enabling technology is recommender systems, which employ all sorts of techniques to predict the preferences of human subjects (e.g. the likes and dislikes towards a movie). A typical system architecture is shown in Figure \ref{system}.


So far, a lot of generic recommender algorithms have been proposed, as surveyed in \cite{BOOK2011}. Recently, deep learning has become a very powerful tool and has been used to numerous applications, including recommender \cite{arXiv2017}. Nevertheless, the collaborative filtering recommender systems are most popular and well-known due to their explainable nature (e.g. you like x so you may also like y). Given a user set $\mathcal{U}=\{1, 2, \cdots, N\}$ and their rating vectors $\mathbf{R}_i$ for $i \in \mathcal{U}$, let $\mathcal{R}$ denote the set of $(i,j)$ such that $r_{i,j} \neq 0$. One of the most popular collaborative filtering algorithms is based on low-dimensional factor models, which derive two feature matrices $\mathbf{U}$ and $\mathbf{V}$ from the rating matrix. The feature vector $\textbf{u}_i$ denotes user $i$'s interest and the feature vector $\textbf{v}_j$ denotes item $j$'s characteristics. Every feature vector has the dimension $k$, which is often a much smaller integer than $M$ and $N$. In implementations, $\mathbf{U}=\{\textbf{u}_i\}_{1 \leq i\leq N}$ and $\mathbf{V}=\{\textbf{v}_j\}_{1 \leq j \leq M}$ are often computed by minimizing the following function:
\begin{equation}\label{eq:RMSE}
  \underset{\mathbf{U},\mathbf{V}}{min}\frac{1}{|\mathcal{R}|}\sum_{(x,j)\in \mathcal{R}}(r_{x,j}- \langle \textbf{u}_x,\textbf{v}_j \rangle)^2+\lambda\sum_{x\in \mathcal{U}}||\textbf{u}_x||^2_2+\mu\sum_{j\in \mathbf{I}}||\textbf{v}_j||^2_2
\end{equation}
for some positive parameters $\lambda, \mu$, typically through the stochastic gradient descent (SGD) method or its variants. Note that one advantage of the latent factor based collaborative filtering is its better resistance to robustness attacks than the neighbourhood-based ones \cite{AAAI2006}.

\subsection{Privacy and Robustness Issues}
\label{sub:properties}

Besides the likes and dislikes, users'preferences might lead to inferences towards other sensitive information about the individuals, e.g. the religion, political orientation, and financial status. When a user is involved in a recommender system with a pseudonym, there is the risk of re-identification. For instance, Weinsberg et al. \cite{RECSYS2012} demonstrated that what has been rated by a user can potentially help an attacker identify this user. Privacy issues have been recognized for a long time and a lot of solutions have been proposed today, as surveyed in \cite{SMR2013,RSH2015}. 
Robustness is about controlling the effect of manipulated inputs, and is a fundamental issue for recommender systems. Its importance can be easily seen from the numerous scandals, including fake book recommendations \footnote{\url{https://tinyurl.com/y9nyo8y9}}, fake phone recommendations \footnote{\url{https://tinyurl.com/ycc8lujh}} and malicious medical recommendations \footnote{\url{https://tinyurl.com/ybuevrwq}}. In their seminal work, Lam and Riedl \cite{WWW2004} investigated the concept of shilling attacks, where a malicious company lies to the recommender system (or, inject fake profiles) to have its own products recommended more often than those from its competitors. Following this, a number of works have been dedicated to the investigation of different robustness attacks and corresponding countermeasures. Interestingly, Sandvig, Mobasher, and Burke \cite{AAAI2006} empirically showed that model-based algorithms are more robust than memory-based algorithms;  Cheng and Hurley \cite{IAAI2009} proposed informed model-based attacks against trust-aware solutions, and demonstrated it against the privacy-preserving solution by Canny \cite{SIGIR2002}.


Clearly, robustness attacks pose a threat to the business perspective of the RecSys and subsequently impact the quality of service for the users. Privacy is increasingly becoming a concern for the privacy-aware users, and it is also a concern for the RecSys when it wants to deploy a \emph{machine learning as a service} business model \cite{USENIX2016}. Unfortunately, privacy and robustness have a \emph{complementary} yet \emph{conflicting} relationship. On the complementary side, it is clear that privacy disclosure can lead to more successful robustness attacks as the attacker can adapt its attack strategy accordingly, and a robust system reduces the attack surface for the privacy attackers who injects fake profiles to infer the honest users' information based on the received outputs. On the conflicting side, a privacy-preserving recommender makes it harder to combat robustness attacks because the robustness attack detection algorithms will not work well when all users' inputs are kept private. We elaborate on this aspect in Section \ref{sec:challenges}.


\subsection{Our Contribution}

In this paper, we aim  at a comprehensive investigation of the privacy and robustness issues for recommender systems, by considering both the  \emph{model training} and the \emph{prediction computing} stages. To this end, we first provide a general system architecture and present a high-level security model accordingly. We then review the existing privacy-preserving latent factor based recommender solutions and identify their potential issues. Particularly, we notice that most cryptographic solutions have mainly aimed at the privacy protection for the \emph{model training} stage without paying much attention to the \emph{prediction computing} stage. This consequently results in serious privacy issues in practice. We also highlight that existing privacy-preserving solutions make it harder to detect and prevent robustness attacks. 

Towards privacy-preserving solutions that respect robustness attack detection, we separately address the issues in the \emph{model training} and \emph{prediction computing} stages. For the former, we show that existing solutions can be adapted, particularly it is straightforward for the expert-based ones such as that from \cite{ArXiv2018}. As to the latter, we propose two new cryptographic protocols, one of which involves an extra proxy. Our experimental results show that both protocols are very efficient with respect to practical datasets. The employed \emph{privacy by design} approach, namely returning unrated items whose approximated predictions are above a threshold, might have profound privacy implications, nevertheless we leave a detailed investigation as future work.

\subsection{Organisation}

The rest of the paper is organised as follows. In Section \ref{sec:model}, we introduce a generic recommender system architecture that consists of two stages: model training and prediction computing. Accordingly, we present a high-level security model. In Section \ref{sec:challenges}, we analyse some representative privacy-preserving recommender solutions and identify their deficiencies in our security model. In Section \ref{sec:framework}, we present a solution framework to demonstrate how to construct secure recommender solutions in our security model. In Section \ref{nothird}, we propose a new privacy-preserving protocol for prediction computing, which does not involve a third-party proxy. In Section \ref{third}, we propose a new privacy-preserving protocol for prediction computing, which is more efficient but relies on a proxy. In Section \ref{sec:con}, we conclude the paper.

\section{System Architecture and Security Model}
\label{sec:model}



\begin{figure}[h]
\centering
\includegraphics[scale=0.52]{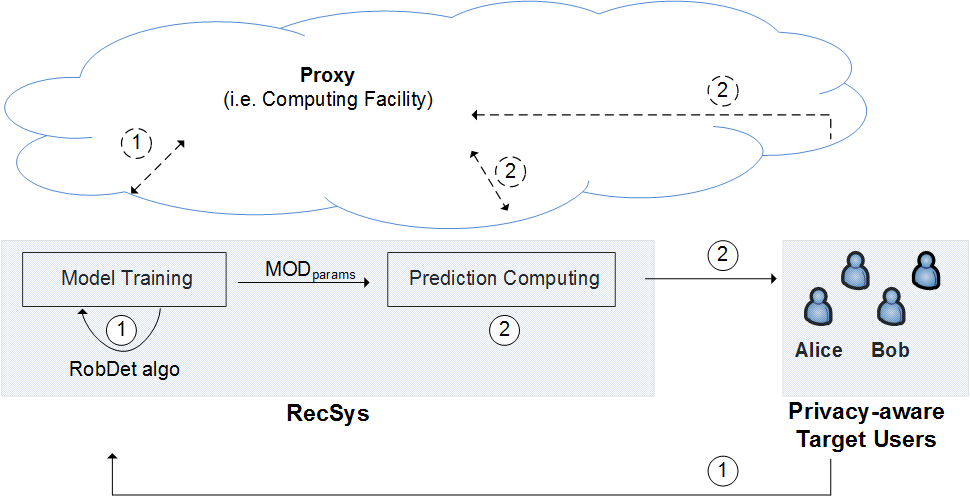}
\caption{Recommender as a Service Architecture}
\label{architecture}
\end{figure}

We assume the RecSys builds recommender models and offers recommendation as a service to the users. If some users do not care about their privacy, then they can offer their rating vectors directly to the RecSys to receive recommendations. In addition, the Recsys may collect as much non-private data as possible in order to build an accurate recommender model. We assume there are privacy-aware users who are not willing to disclose their rating vectors while still wishing to receive recommendations. Our main objective is to design solutions to guarantee that, from the view point of a privacy-aware user Alice,

\begin{itemize}
\item She receives high-quality recommendations, by avoiding the robustness attacks mentioned in Section \ref{sub:properties}.
\item She minimizes the information disclosure about her rating vector, under the prerequisite that she receives high-quality recommendations.
\end{itemize}

For our recommender as a service, we assume a system architecture shown in Figure \ref{architecture}. We note that existing collaborative filtering recommender systems typically have the \emph{model training} and \emph{prediction computing} stages even though they might not mention them explicitly. In addition, it is also quite often that a proxy (i.e. cloud computing facility) is employed to carry out the massive computations (e.g. Netflix heavily uses Amazon cloud services). It is worth emphasizing that many privacy-preserving solutions (particularly cryptographic solutions) also introduce such a third party, e.g. the crypto service provider in the term of \cite{ASIACCS2016} and \cite{CCS2013}. For the different usage scenarios, the trust assumptions on the proxy can vary a lot, and we elaborate on it later. Next, we briefly introduce what will happen in the two stages.

\begin{enumerate}
\item In the \emph{model training} stage, labeled in Figure \ref{architecture}, the RecSys trains a model, e.g. similarities between items (or users) in neighbourhood-based recommenders and feature matrices for users and items in latent model based ones, based on data from one or more sources. To clean the data and detect robustness attacks, before the training, we suppose that the RecSys will run an algorithm $\mathsf{RobDet}$ over the training dataset. To simplify our discussion, we assume the the output of $\mathsf{RobDet}$ is a binary bit for every input profile (i.e. rating vector). If it is 0, then the profile is deemed as malicious so that will not be used in the training.

\item After training, we refer to the output of the model training stage as a set of parameters $MOD_{params}$. Note that the parameters might be in an encrypted form when privacy protection has been applied. In the \emph{prediction computing} stage, the RecSys uses the model parameters $MOD_{params}$ and possibly Alice's rating vector to infer Alice's preferences.
\end{enumerate}

\subsection{The Proposed Security Model (high level)}
\label{subsec:model}

We make the following general assumptions related to security. First of all, we assume the communication channel is secured with respect to confidentiality and integrity in the sense: (1) an honest user can be assured that his input will reach the RecSys or another intended party without being eavesdropped on and manipulated; (2) the RecSys can be assured that the honest user, who initiates the communication, will receive the message without being eavesdropped on and manipulated. It is worth stressing that there is no guarantee whether  RecSys knows the true identity of the user it is communicating with. Secondly, we assume that the RecSys is a rational player and offers recommendation as a service and a user offers monetary rewards for receiving recommendations. Without this assumption, there will not be any guarantee for achieving privacy and robustness because the RecSys will deviate from the protocol for any possible benefits.

Regarding robustness, we require that the RecSys is able to (efficiently) run any chosen $\mathsf{RobDet}$ algorithm over the training dataset to identify the malicious profiles, i.e. rating vectors, as we have described in the beginning of this section. The output of $\mathsf{RobDet}$ should be the same regardless what privacy protection mechanisms have been deployed.

Regarding privacy, we consider the following specific requirements. Note that, similar to the semantic security of encryption schemes, indistinguishability-based games can be defined to formally capture all requirements. We skip the details here, partially due to the fact that the cryptographic primitives we use (e.g. homomorphic encryption) guarantee indistinguishability straightforwardly.

\begin{itemize}
\item \emph{Alice's privacy against RecSys.} If the RecSys does not collude with the proxy, then it learns nothing about Alice's input and output except for information implied in the output of $\mathsf{RobDet}$ (i.e. whether or not Alice's profile is suspicious if it has been used in the model training stage).

\item \emph{Alice's privacy against Proxy.} If the proxy does not collude with the RecSys, then it learns nothing about Alice's input and output.

\item \emph{Alice's privacy against other users.} Other users do not learn more information about Alice's rating vector than that implied in the legitimate outputs they receive.

\item \emph{RecSys's privacy against Alice and other users.} Alice and other users do not learn more information than that implied in the legitimate outputs they receive.
\end{itemize}

As a remark, in many existing solutions reviewed in Section \ref{sub:crypto}, the legitimate outputs can contain too much private information. This has motivated our privacy-by-design approach in Section \ref{privacy-by-design}. As an informal requirement, when both the RecSys and the Proxy are compromised simultaneously, the information leakage about the privacy-aware users' data should also be minimized. To this end, we note that most existing solutions except for the expert-based ones will leak everything.

\section{Literature Work and Standing Challenges}
\label{sec:challenges}


Regardless efficiency, designing a secure recommender system is a very challenging task. For example, applying statistical disclosure mechanisms does not guarantee security, as Zhang et al. \cite{SDM2006} showed how to recover perturbed ratings in the solutions by Polat and Du \cite{ICDM2003}. Employing advanced cryptographic primitives is also not a panacea, as Tang and Wang \cite{ESORICS2015} pointed out a vulnerability in the homomorphic encryption based solution by Jeckmans et al. \cite{ESORICS2014}. Next, we analyse some representative solutions from the literature and identify the standing challenges.

\subsection{Preliminary on Building Blocks}

We use the notation $x \getsr Y$ to denote that $x$ is chosen from the set $Y$ uniformly at random. A public key encryption scheme consists of three algorithms $(\mathsf{Keygen}, \mathsf{Enc}, \mathsf{Dec})$: $\mathsf{Keygen}(\lambda, L)$ generates a key pair $(PK,SK)$; $\mathsf{Enc}(m, PK)$ outputs a ciphertext $c$; $\mathsf{Dec}(c, SK)$ outputs a plaintext $m$. Some schemes, e.g. Paillier \cite{EUROCRYPT1999}, are additively homomorphic, which means there is an operator $\oplus$ such that $\mathsf{Enc}(m_1, PK) \oplus \mathsf{Enc}(m_2, PK)= \mathsf{Enc}(m_1+m_2, PK)$. While some recent somewhat homomorphic encryption (SWHE) schemes are both additively and multiplicatively homomorphic to a certain number of operations, which means there are operators $\oplus$ and $\otimes$ such that $\mathsf{Enc}(m_1, PK) \oplus \mathsf{Enc}(m_2, PK)= \mathsf{Enc}(m_1+m_2, PK)$ and $\mathsf{Enc}(m_1, PK) \otimes \mathsf{Enc}(m_2, PK)= \mathsf{Enc}(m_1m_2, PK)$. In practice, one of the most widely-used SWHE library is Simple Encrypted Arithmetic Library (SEAL) from Microsoft \cite{IEEE2017}, which is an optimized implementation of the YASHE scheme \cite{IMA2013}. Note that homomorphic subtraction $\ominus$ can be directly defined based on $\oplus$ and with similar computational cost.

\subsection{Examining some Cryptographic Solutions}
\label{sub:crypto}

Cryptographic solutions aim at minimizing the information leakage in the computation process, by treating the recommender as a large-scale multi-party computation protocol. When designing privacy-preserving solutions, it has become a common practice to introduce one or multiple third parties, not all of which are supposed to collude with each other, in order to eliminate a single trusted third party and improve efficiency. Nikolaenko et al. \cite{CCS2013} and Kim et al. \cite{ASIACCS2016} introduced a CSP (i.e. crypto service provider) and employed garbled circuits and homomorphic encryption respectively to perform privacy-preserving matrix factorization.
These solutions put the emphasis on the matrix factorization step (i.e. model training stage) while failing to pay more attention to the prediction computing stage. In \cite{CCS2013}, it is proposed that every user $i$ is given his own feature vector $\textbf{u}_i$ so that it can interact with the RecSys and CSP to retrieve predictions on all items (i.e. $\textbf{u}_i(\textbf{v}_i)^T$ $(1 \leq j \leq M)$). In reality, the users do not need to know his feature vector and the predictions to all items, they only need to know the items they might like. In more detail, there are several concerns.
    \begin{itemize}
    \item Given the fact that $M << N$ (i.e. the number of items are far less than the user population), a small number of colluded users can recover the item feature matrix $\mathbf{V}$, based on which they can try to infer information about the rest of the population. This leads to unnecessary information leakages against the honest users.
    \item The malicious users might make illegal use of the recovered $\mathbf{V}$, through providing recommendation services using technologies, such as incremental matrix factorization. Besides the potential privacy concern, this may hurt the business model of the RecSys.
    \item Privacy-preserving mechanisms, such as encryption and garbled circuits, make it very difficult to detect Sybil attacks, where an attacker injects fake profiles into the system and then it can (1) try to infer private information based on the outputs to these fake profiles (2) and mount robustness attacks. Canny \cite{SP2002} used zero-knowledge proof technique to fight against ill-formed profiles (i.e. ratings set beyond $\{0,\cdots,5\}$), but it is not effective against Sybil attacks. With respect to the robustness attacks in reality, the forged rating vectors are always well-formed (but the rating values in these forged rating vectors follow maliciously defined distributions), otherwise the RecSys can easily identify the ill-formed ones in plaintext. To detect and prevent robustness attacks, special detection algorithms need to be executed on the input rating vectors in the privacy-preserving solutions.
    \end{itemize}


When training a recommender model, it is unnecessary to always take into the ratings from all possible users. Amatriain et al. \cite{SIGIR2009} introduced recommender system based on expert opinions, and showed that the recommendation accuracy can be reasonably good even if a target user's data is not used in training the model. Following this concept, Ahn and Amatriain \cite{WI2010} proposed a privacy-preserving distributed recommender system, and similar concept has been adopted in \cite{SCC2017,ArXiv2018}. The solution from  \cite{ArXiv2018} is very interesting because it leads to very efficient solutions. We briefly summarize it below.
\begin{itemize}
\item In the model training stage, suppose the expert data set consists of rating vectors $\mathbf{R}_t$ $(1 \leq t \leq N)$. The model parameters are denoted as $\Theta = \{\textbf{A},\textbf{Q}, \textbf{b}_t^*, \textbf{b}_j^* \}$, where $\textbf{A}$ and $\textbf{Q}=\{\textbf{q}_1, \cdots, \textbf{q}_M\}$ are two independent item feature spaces, $\mu$ is the global rating average, $b_t = \bar{r}_t - \mu$ and $b_j = \bar{r}_j - \mu$ are the average rating for user $t$ and item $j$ respectively, $\textbf{b}_t^* = \{b_t^* \}_{t=1}^N$ , $\textbf{b}_j^* = \{b_j^* \}_{j=1}^M$ are the user and item bias vectors. Suppose user $t$ has not rated item $j$, his preference is formulated as follows.
    \begin{equation}
    \hat{r}_{t,j} = \mu+ b_t +b_j+b_t^*+b_j^* + (\mathbf{R}_t\textbf{A})\textbf{q}_j^T
    \end{equation}
    Similar to other solutions, SGD can be used to learn the parameter $\Theta$.
\item In the prediction computing stage, suppose the user $i$ is not in the expert dataset and has the rating vector $\mathbf{R}_i$ and rating average $\overline{\mathbf{R}_i}$, the prediction for rating $r_{i,j}$ is computed as follows
    \begin{equation}
    \label{pred}
    \hat{r}_{i,j} = \overline{\mathbf{R}_i} + b_j+ \frac{\sum_{t=1}^{N}b_t^*}{N} +b_j^* + (\mathbf{R}_i \textbf{A})\textbf{q}_j^T
    \end{equation}
\end{itemize}
According to their experimental results, the accuracy of the predictions are almost the same to the state-of-the-art recommender systems even though  user $i$ is not required to be involved in the model training stage. As such, Wang et al. \cite{ArXiv2018} further proposed an efficient privacy-preserving protocol based on Paillier encryption scheme, so that the prediction (i.e. Equation (\ref{pred})) can be computed in the encrypted form. Unfortunately, their solution allows a malicious user $i$ or several of such users to straightforwardly recover $b_j + \frac{\sum_{t=1}^{N}b_t^*}{N} +b_j^*, \textbf{A}\textbf{q}_j^T$ $(1 \leq j \leq M)$, which are functionally equivalent to the model parameters $\Theta$, by solving some simple linear equations. This attack poses severe threats against the \emph{recommendation as a service} objective and the privacy of the RecSys, claimed in \cite{ArXiv2018}.



\subsection{Examining the DP-based Solutions}

While cryptographic solutions might provide provable security for the computation, they do not consider the information leakages from the legitimate outputs. In particular, the inference against an honest user or a group of honest users might be very severe when the attacker has effectively controlled part of the population (e.g. by launching Sybil attacks). Following the seminal work of McSherry and Mironov \cite{SIGKDD2009}, researchers have tried to apply the differential privacy concept to the prevent information leakages from recommender outputs, e.g. \cite{RECSYS2015a,ICALP2006,TCC2006}.

One of the main issues with DP-based approach is how to set the privacy parameter $\epsilon$. Specific to recommender systems, it is unrealistic to predefine a privacy budget, because the recommender algorithm (i.e. model training stage) will be executed hundreds, thousands or more times. With respect to the sequential composition theorem, the privacy guarantee becomes $N\cdot \epsilon$ after $N$ executions of the recommender algorithm.  In this case, to maintain a meaningful level of privacy protection, the privacy parameter $\epsilon$ in every execution needs to be so small such that the recommendation accuracy will be totally destroyed. Besides, most DP-solutions assume a trusted curator (e.g. RecSys), which means there is no privacy against this party. In other solutions (e.g. local differential privacy\cite{CCS2016}) no trusted curator is required but it will severely interfere with robustness attack detection operations. For example, a privacy-aware user who prefers a higher level of privacy protection might be prone to be classified as malicious due to the extensive perturbation of his rating vector.

\section{Modular Solution Constructions}
\label{sec:framework}

In this section, we present modular solutions which secure both the model training and prediction computing stages. We first introduce a privacy-by-design concept to minimize information leakages from the outputs, and then describe two types of constructions. In one type of construction, the RecSys trains its recommender model without relying on privacy-aware users' data, while in the other the RecSys needs privacy-aware users' data to train the model so that these users can receive meaningful recommendations. For notation purpose, we refer to them as \emph{Expert-based Solution} and \emph{Self-based Solution} respectively.

Note that for the constructions, we leave the detailed description of privacy-preserving protocols for the prediction computing stage to Section \ref{nothird} and \ref{third}.

\subsection{\emph{Privacy by Design} Concept}
\label{privacy-by-design}

\begin{figure}[h]
\centering
\includegraphics[scale=0.18]{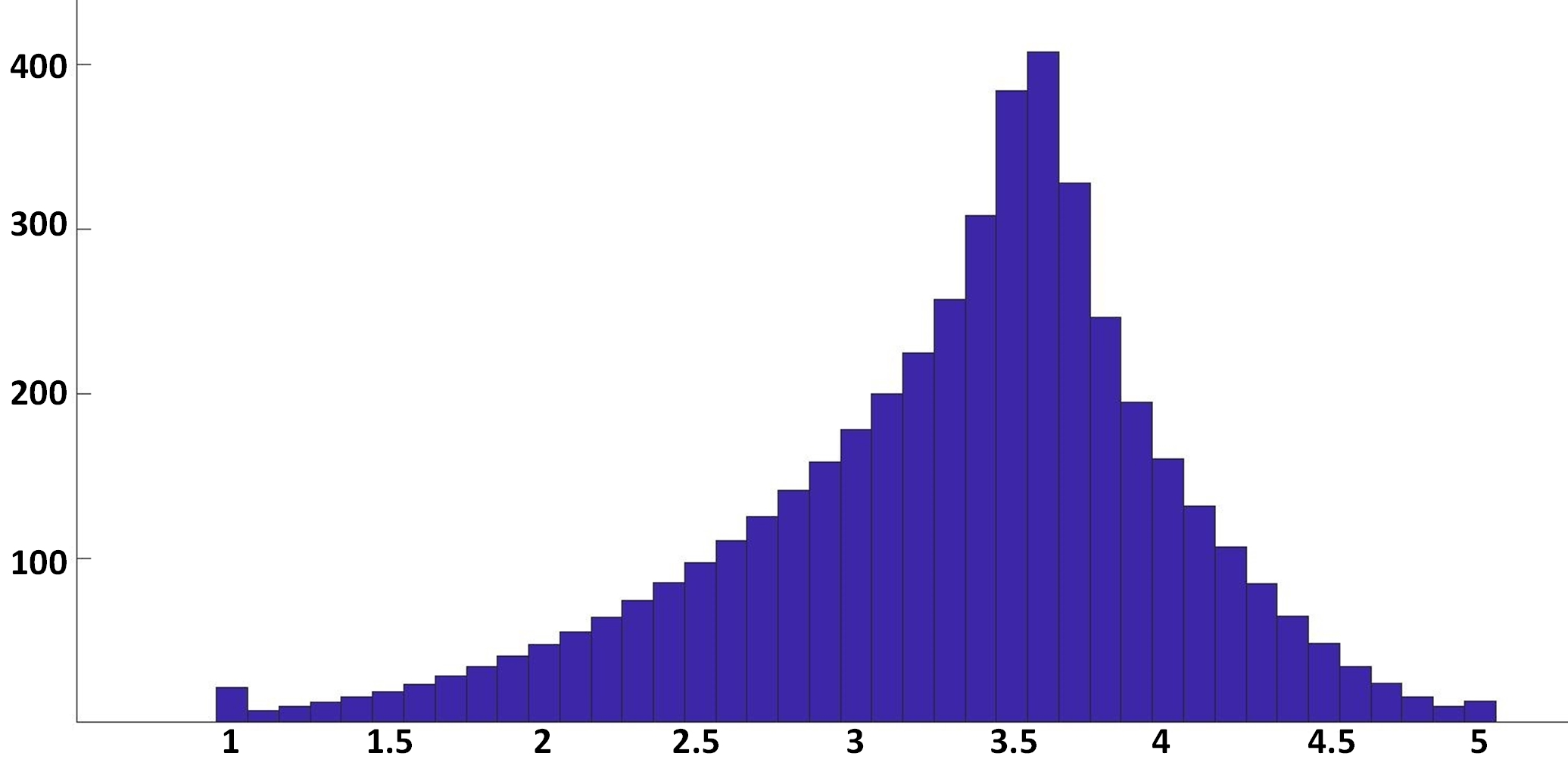}
\caption{Prediction Distribution}
\label{resolution}
\end{figure}

In Section \ref{sub:crypto}, we have shown that the legitimate outputs in the solutions from \cite{CCS2013,ArXiv2018} contain a lot of unnecessary information and can leak the recommender model to a small group of malicious users. To avoid such problems, we enforce the \emph{privacy by design} concept by restricting the output to any user to be the unrated items whose predictions are above a threshold in the proposed prediction computing stage. This significantly reduces the leakage of $MOD_{params}$ to the user and also allow more efficient protocol design. In reality, the predictions to many items can be quite close so that it is very subtle to only return Top-k (say k=20) items. For example, for MovieLens 1M Dataset with 1 million ratings from 6000 users on 4000 movies\footnote{\url{https://grouplens.org/datasets/movielens/1m/}}, the distribution of predictions is shown in Figure \ref{resolution}, where the horizontal Axis stands for the prediction value and vertical Axis stands for the number of predictions that possess the value. Note that all predicted ratings have been rounded to have one decimal place. Intuitively, as an example, it makes more sense to return the unrated items whose ratings are 4.9 or 5. Put it another way, we only need to return the items whose predicted ratings fall into the set $\{V_1=5, V_2=4.9\}$.


\subsection{Privacy-preserving and Robust Expert-based Solution}
\label{subsec:expert}

In this solution, we adopt the recommender algorithm \cite{ArXiv2018}, which has the nice property that the privacy-aware user Alice does not need to share her rating vector with the RecSys to train the recommender model and the process of model training is very simple. Note that in some other expert-based recommender systems, Alice's data may not be needed to train the model but the process of model training will be much more complex (i.e. often retraining the recommender model is required before being able to generate recommendations for Alice).

\begin{enumerate}
\item In the solution, the \emph{model training} stage is very straightforward. Given an expert dataset, the RecSys can first run any robustness attack detection algorithm $\mathsf{RobDet}$ to figure out the outliers or even malicious profiles. Then, the RecSys can learn the model parameters $\Theta = \{\textbf{A},\textbf{Q}, \textbf{b}_t^*, \textbf{b}_j^* \}$ from the expert dataset, which is publicly available to the RecSys. More information can be seen from Section \ref{sub:crypto}.

\item Let's assume that Alice is labelled as user $i$ in the privacy-aware user group, the \emph{prediction computing} stage consists of the following steps.

    \begin{enumerate}
    \item User $i$ generates a public/private key pair $(pk_i, sk_i)$ for an SWHE scheme, and shares the public key $pk_i$ with RecSys.

    \item User $i$ sends $\llbracket \mathbf{R}_i \rrbracket_{pk_i}$ and $\llbracket \overline{\mathbf{R}_i} \rrbracket_{pk_i}$ to the RecSys, which may require the user to prove that the encrypted $\mathbf{R}_i$ is well formed similar to what has been done in \cite{SP2002}.

    \item If everything is ok, the RecSys can predict user $i$'s preference on item $j$ as
    \begin{equation}
    \llbracket \hat{r}_{i,j} \rrbracket_{pk_i}  = \llbracket \overline{\mathbf{R}_i} \rrbracket_{pk_i} \oplus b_j \oplus \frac{\sum_{t=1}^{N}b_t^*}{N} +b_j^* \oplus \llbracket \mathbf{R}_i \rrbracket_{pk_i} \textbf{A}\textbf{q}_j^T
    \end{equation}

    \item If there is no proxy, user $i$ and the RecSys run the protocol from Section \ref{nothird} to generate recommendations for user $i$. Otherwise, they run the protocol from Section \ref{third}.
    \end{enumerate}
\end{enumerate}

\subsection{Privacy-preserving and Robust Self-based Solution}
\label{subsec:self}

In this solution, we build on top of the privacy-preserving solutions from \cite{CCS2013} and \cite{ASIACCS2016}.

\begin{enumerate}
\item In the model training stage, we need to augment existing privacy-preserving protocols for the model training stage, e.g. those from \cite{CCS2013} and \cite{ASIACCS2016}, to enable privacy-preserving robustness attack detection.
    \begin{itemize}
    \item In case of \cite{CCS2013}, we need to devise a larger garbled circuit, which first evaluates $\mathsf{RobDet}$ and then chooses the unsuspicious inputs to proceed with the matrix factorization procedure.
    \item In case of \cite{ASIACCS2016}, we need to devise a cryptographic protocol that can evaluate $\mathsf{RobDet}$ algorithm on the same encrypted inputs to those used in the HE-based matrix factorization algorithm.
    \end{itemize}
    A seamless augmentation will depend on the specific robustness attack detection algorithms, so that in this paper we skip the details, which can be an interesting future though.

\item At the end of the privacy-preserving matrix factorization, either from \cite{CCS2013} or \cite{ASIACCS2016}, the RecSys will possess $\llbracket\textbf{u}_i \rrbracket_{pk_c}$ $(1 \leq i \leq N)$ and $\llbracket\textbf{v}_i \rrbracket_{pk_c}$ $(1 \leq j \leq M)$, where $(PK_c, SK_c)$ is an SWHE public/private key pair from the CSP (or Proxy in our system structure). The participants (i.e. user $i$, RecSys, and Proxy) then perform the following steps.
    \begin{enumerate}
    \item The RecSys computes user $i$'s preference on item $j$ as $\llbracket \hat{r}_{i,j} \rrbracket_{pk_c}=\llbracket \textbf{u}_i \rrbracket_{pk_c} \otimes \llbracket \textbf{v}_j \rrbracket_{pk_c}$ for every $1 \leq j \leq M$.
    \item For every $1 \leq j \leq M$, the RecSys selects a random number $r_j$, then computes and sends $\llbracket \hat{r}_{i,j} \oplus r_j \rrbracket_{pk_c}$ to the proxy. User $i$ generates a Paillier public/private key pair $(PK_i, SK_i)$ and send $PK_i$ to the Proxy.
    \item The Proxy decrypts $\llbracket \hat{r}_{i,j} \oplus r_j \rrbracket_{pk_c}$ and re-encrypts the plaintext to obtain $\llbracket \hat{r}_{i,j} \oplus r_j \rrbracket_{pk_i}$ $(1 \leq j \leq M)$.
    \item For every $1 \leq j \leq M$, the RecSys removes $r_j$ from $\llbracket \hat{r}_{i,j} \oplus r_j \rrbracket_{pk_i}$ to obtain $\llbracket \hat{r}_{i,j} \rrbracket_{pk_i}$.
    \end{enumerate}

\item If there is no proxy, user $i$ and the RecSys run the protocol from Section \ref{nothird} to generate recommendations for user $i$. Otherwise, they run the protocol from Section \ref{third}.

\end{enumerate}

It is clear that the model training stage of our expert-based solution satisfies all our robustness and privacy expectations, while the privacy analysis depends on the protocols from Section \ref{nothird} and \ref{third} because the existing steps do not leak information due to the encrypted operations and randomization. For the self-based solution, we can guarantee the same level of privacy and robustness protection, although it will apparently be less efficient than the previous expert-based one.

\section{Privacy-preserving Prediction Computing}
\label{nothird}

In this section, we describe a privacy-preserving protocol for user $i$ to learn the unrated items whose predictions fall into a set $\{V_1, \cdots V_T\}$, without relying on a proxy. Here $T$ will be a small integer, which may be 2 or 3 in practice referring to the example in the previous section. Observing that privacy-preserving protocols for the model training stage often output integer predictions (in encrypted form), because they need to scale the intermediary computation results in order to be compatible with the cryptographic tools such as homomorphic encryption algorithms. Therefore, we assume the RecSys possesses the encrypted predictions $\llbracket x_j \cdot \theta + y_j \rrbracket_{pk_i}$ for every $1 \leq j \leq M$ at the end of the privacy-preserving model training stage. We explicitly present the ratings according to a unit $\theta$, because in our protocol the recommendations will only be based on the $x_j$ part and the $y_j$ part is rounding off.

\subsection{Description of the Proposed Protocol}

At the beginning of the prediction computing stage, we suppose user $i$ possesses two public/private key pairs: one is $(PK_i, SK_i)$ for the Paillier scheme which has been setup in Section \ref{subsec:expert} and \ref{subsec:self} while the other is new key pair $(PK_i', SK_i')$ for a SWHE encryption scheme \cite{seal}. The public keys $PK_i, PK_i'$ are shared with the RecSys. As shown in Figure \ref{fig:nothird}, the protocol runs in two phases where $\lambda$ is the security parameter.

\begin{figure}[h]
\begin{scriptsize}
\begin{displaymath}
\begin{array}{l|lcr}
\hline
&\mbox{\bf User $i$}  & & \mbox{\bf RecSys}  \\
&Paillier: (PK_i, SK_i)&& \llbracket x_j \cdot \theta + y_j \rrbracket_{pk_i} \\
&SWHE: (PK_i', SK_i')&& (1 \leq j \leq M)\\
\hline
\multirow{8}{*}{ \parbox{1.7cm}{\emph{\bf{Reduction}}\\ \\ \\ for each \\ $1 \leq j \leq M$}}
&&& r_{j1} \getsr \{0,1\}^{\lambda}\\
&&& r_{j2} \getsr [0, \theta)\\
&&& \Delta_j=\llbracket x_j \cdot \theta + y_j \rrbracket_{pk_i} \oplus (r_{j1} \cdot \theta + r_{j2})\\
&&&  =\llbracket (x_j+r_{j1}) \cdot \theta + y_j +r_{j2} \rrbracket_{pk_i} \hspace{1cm}\\
&&&\\
&&\xleftarrow{\Delta_j}&\\
&&&\\
&\alpha_j=\mathbf{Dec}(\Delta_j, SK_i) &&\\
&  \hspace{0.44cm}    =(x_j+r_{j1}) \cdot \theta + y_j +r_{j2} &&\\
&\beta_j=\frac{\alpha_j - (\alpha_j \mod \theta)}{\theta} &&\\
&    \hspace{0.44cm} =x_j + r_{j1} + \epsilon_j  &&\\
&\mbox{If item $j$ is unrated}: &&\\
&\hspace{0.3cm}\Gamma_j = \mathsf{Enc}(\beta_j, PK_i')&&\\
&\mbox{Otherwise}: &&\\
&\hspace{0.3cm} r_{j3} \getsr \{0,1\}^{\lambda}&&\\
&\hspace{0.3cm} \Gamma_j = \mathsf{Enc}(r_{j3}, PK_i')&&\\
&&\xrightarrow{\Gamma_j}&\\
&&& \Phi_j = \Gamma_j \ominus r_{j1} \\
\hdashline
\multirow{6}{*}{ \parbox{1.7cm}{\emph{\bf{Evaluation}}}}
&&& \displaystyle \Omega_j = \prod_{x=1}^{T}(\Phi_j \ominus V_x)\\
&&& \displaystyle \Psi_j = \mathsf{RAND}(\Omega_j)\\
&&\xleftarrow{\Psi_j (1 \leq j \leq M)}&\\
&\{j| \mathsf{Dec}(\Psi_j, SK_i')\neq 0\} &&\\
\hline
\end{array}
\end{displaymath}
\caption{Learning Membership in $\{V_1, \cdots V_T\}$ without a Proxy}
\label{fig:nothird}
\end{scriptsize}
\end{figure}


In the \emph{reduction} phase, the RecSys and user $i$ round off the $y_j$ part in the encrypted predictions. Specifically, for every $1 \leq j \leq M$, the following operations will be carried out.
\begin{enumerate}
\item The RecSys first randomizes $x_j$ and $y_j$ to generates $\Delta_j$ for user $i$.

\item Then, user $i$ obtains the randomized prediction value $\alpha_j$ through decryption and then computes $\beta_j$, which is the randomized $x_j$ in an approximation form with $\epsilon_j \in \{0,1\}$. Finally, user $i$ encrypts $\beta_j$ under his own SWHE public key if item $j$ is unrated, and encrypts a random value otherwise.

\item After receiving $\Gamma_j$, the RecSys homomorphically removes the randomization noise $r_{j1}$ to obtain $\Phi_j$, which is a ciphertext for $x_j + \epsilon_j$ if item $j$ is unrated and a ciphertext for a random value otherwise.
\end{enumerate}
In the \emph{Evaluation} phase, for every $1 \leq j \leq M$, the RecSys computes $\Omega_j$ through $T$ homomorphic subtractions and $T-1$ multiplications, which is a ciphertext for $0$ if the plaintext corresponding to $\Phi_j$ falls into $\{V_1, \cdots V_T\}$ and a ciphertext for a non-zero value otherwise. In order to hide the non-zero values, the RecSys randomize $\Omega_j$ via the $\mathsf{RAND}$ function, e.g. homomorphicly multipling a random number, to obtain $\Psi_j$, which can be decrypted by user $i$ to learn the index of recommended items.

\subsection{Security and Performance Analysis}
\label{nothirdanalysis}

The operations in the protocol are done with encrypted data and randomization has been applied to the predictions revealed to user $i$. As such the protocol only reveals the desired items to user $i$ while leaks nothing to the RecSys.

For Paillier, we set the size of $N$ to be 2048, and for SWHE we use Microsoft SEAL library. We select the ciphertext modulus $q=2^{226}-2^{26}+1$, the polynomial modulus $p(x)=x^{8192}+1$. Using Chinese Reminder Theorem, we select two 40-bit primes to represent the plaintext space of $2^{80}$. The primes are 1099511922689 and 1099512004609. By packing 8192 plaintexts into one ciphertext, we can process 8192 multiplications in one homomorphic multiplication. Based on an Intel(R) Core(TM) i7-5600U CPU \@ 2.60GHz, 8GB RAM, the timing is summarized in Table \ref{tab:impcost}.

\vspace{-0.2cm}

    \begin{table}[h]
        \centering
        \begin{tabular}{|c|c|c|c|}
        \hline
             $\mathsf{Paillier.Enc}$ &$\mathsf{Paillier.Dec}$&$\mathsf{Paillier}.\oplus$&$\mathsf{SWHE.Enc}$\\
            \hline
               31.30 ms & 12.88 ms&8.50 $\mu$s&52.43 ms \\
        \hline
             $\mathsf{SWHE.Dec}$  & $\mathsf{SWHE}.\otimes$ & partial $\mathsf{SWHE}.\otimes$ & $\mathsf{SWHE}.\oplus$ \\
            \hline
               39.63 ms & 207.76 ms & 70.28 ms& 742 $\mu$s\\
        \hline
        \end{tabular}
        \caption{Costs for SWHE and Paillier}
        \label{tab:impcost}
    \end{table}

\vspace{-0.2cm}

The number of different cryptographic operations for the proposed protocol are summarized in Table \ref{tab:comwithoutttp}.  In the last column, we estimate the real-world running time based on the aforementioned benchmarking results, where $M=4000$ by assuming the MovieLens 1M Dataset and $T=2$. Note that this dataset has been used in Section \ref{privacy-by-design}.

\vspace{-0.2cm}

    \begin{table}[h]
        \centering
        \begin{scriptsize}
        \begin{tabular}{|c|c|c|c|c|c|c|c||c|}
        \hline
              &$\mathsf{Paillier.Dec}$  & $\mathsf{Paillier}.\oplus$ & $\mathsf{SWHE.Enc}$ &$\mathsf{SWHE.Dec}$  & $\mathsf{SWHE}.\otimes$ & partial $\mathsf{SWHE}.\otimes$ & $\mathsf{SWHE}.\oplus$ & {\bf Time}\\
        \hline
             User &  $M$ & & $M$ & $M$ & & & & 420 s\\
        \hline
             RecSys &  & $M$ & & $M$ & $M(T-1)$& & $M(T+1)$& 998 s\\
        \hline
        \end{tabular}
        \end{scriptsize}
        \caption{Computational Complexities}
        \label{tab:comwithoutttp}
    \end{table}

\vspace{-0.2cm}

With respect to the MovieLens 1M Dataset, we consider the standard case where user $i$ and the RecSys interactively rank the predictions and the RecSys returns the top-ranked items. In order to rank, user $i$ and RecSys need to perform a comparison for two predictions for the RecSys to learn the order of them. Based on the same computer as above, for a comparison with the protocol from \cite{NDSS2015}, the computation time for user $i$ and the Recsys is 175.88 ms and 184.60 ms respectively. Suppose we adopt a standard sorting algorithm to realise the ranking, and the average computation time for the user and RecSys will be 8442.24 s and 8860.80 s, respectively. The time delay due to the communication is about 4800 s, by assuming each computation takes up to 100 ms as in \cite{NDSS2015}. It is clear that our protocol is much more efficient.


\section{Privacy-preserving Prediction Computing with Proxy}
\label{third}

In this section, we describe the protocol that relies on a proxy, and also provide corresponding analysis.

\subsection{Description of the Proposed Protocol}

To enable the new protocol, we make use of a key-homomorphic pseudorandom function $\mathsf{Prf}$ \cite{CRYPTO2014a}. Given $\mathsf{Prf}(k_1, m)$ and $\mathsf{Prf}(k_2, m)$, anybody can compute $\mathsf{Prf}(k_1+k_2, m)=\mathsf{Prf}(k_1, m) \oplus \mathsf{Prf}(k_2, m)$. We describe the two phases in Figures \ref{fig:third-red} and \ref{fig:third-eva}, respectively. As before, $\lambda$ is the security parameter.

\vspace{-0.1cm}

\begin{figure}[H]
\begin{scriptsize}
\begin{displaymath}
\begin{array}{l|lcr}
\hline
&\mbox{\bf User $i$}  & & \mbox{\bf RecSys}  \\
&Paillier: (PK_i, SK_i)&& \llbracket x_j \cdot \theta + y_j \rrbracket_{pk_i} \\
&SWHE: (PK_i', SK_i')&& (1 \leq j \leq M)\\
\hline
\multirow{8}{*}{ \parbox{1.7cm}{\emph{\bf{Reduction}}\\ \\ \\ for each \\ $1 \leq j \leq M$}}
&&& r_{j1} \getsr \{0,1\}^{\lambda}\\
&&& r_{j2} \getsr [0, \theta)\\
&&& \Delta_j=\llbracket x_j \cdot \theta + y_j \rrbracket_{pk_i} \ominus (r_{j1} \cdot \theta - r_{j2})\\
&&&  =\llbracket (x_j-r_{j1}) \cdot \theta + y_j +r_{j2} \rrbracket_{pk_i} \hspace{1cm}\\
&&&\\
&&\xleftarrow{\Delta_j}&\\
&&&\\
&\alpha_j=\mathbf{Dec}(\Delta_j, SK_i) &&\\
&  \hspace{0.44cm}    =(x_j-r_{j1}) \cdot \theta + y_j +r_{j2} &&\\
&\beta_j=\frac{\alpha_j - (\alpha_j \mod \theta)}{\theta} &&\\
&    \hspace{0.44cm} =x_j - r_{j1} + \epsilon_j  &&\\
&r_{j3} \getsr \{0,1\}^{2\lambda} &&\\
&\mbox{If item unrated}: &&\\
&\gamma_j = r_{j3} + \beta_j&&\\
&    \hspace{0.44cm} =x_j - r_{j1} + \epsilon_j +r_{j3}  &&\\
&\mbox{If item rated}: &&\\
&\gamma_j = r_{j3}&&\\
&&\xrightarrow{\gamma_j}&\\
&&&\gamma_j = \gamma_j+r_{j1}\\
\hline
\end{array}
\end{displaymath}
\caption{Reduction Phase}
\label{fig:third-red}
\end{scriptsize}
\end{figure}

\vspace{-0.1cm}

Similar to the case shown in Figure \ref{fig:nothird}, in the \emph{reduction} phase, the RecSys and user $i$ interactively round off the $y_j$ part in the predictions for every $j$. The main difference (and simplification) is that, at the end of the protocol, the RecSys possesses $\gamma_j = x_j +  \epsilon_j +r_{j3}$ if item $j$ has been rated and $\gamma_j = r_{j3}+r_{j1}$ otherwise, while user $i$ possesses the random number $r_{j3}$.

\vspace{-0.1cm}

\begin{figure}[H]
\begin{scriptsize}
\begin{displaymath}
\begin{array}{lcccl}
\hline
\mbox{\bf User $i$} & & \mbox{\bf Proxy} && \mbox{\bf RecSys}\\
\:\: r_{j3} (1 \leq j \leq M)&&&& \gamma_j (1 \leq j \leq M)\\
\hline
&&&&\\
R_j, \mathsf{PM}_j (1 \leq j \leq M) &&&& R_j, \mathsf{PM}_j (1 \leq j \leq M)\\
\mathsf{PM}, \mathsf{H} &&&& \mathsf{PM}, \mathsf{H}\\
K_j (1 \leq j \leq M) \gets \{0,1\}^{3\lambda}&&&&\\
\Upsilon_j =\mathsf{Prf}(K_j-r_{j3}, R_j) &&&&\Theta_j =\mathsf{Prf}(\gamma_j, R_j)\\
(1 \leq j \leq M) &&&&(1 \leq j \leq M)\\
&\xrightarrow{\mathsf{PM}\{(\Upsilon_1, \cdots, \Upsilon_M)\}}&&\xleftarrow{\mathsf{PM}\{(\Theta_1, \cdots, \Theta_M)\}}&\\
&&\Xi &&\\
\Omega_{x,j} =\mathsf{H}(\mathsf{Prf}(K_j+V_x, R_j)) &&&&\\
(1 \leq x \leq T,1 \leq j \leq M)&&&&\\
\Omega_{*, j} = \mathsf{PM}_j\{(\Omega_{1,j}, \cdots, \Omega_{T,j})\}&&&&\\
&\xrightarrow{\mathsf{PM}\{(\Omega_{*,1}, \cdots, \Omega_{*,M})\}}&&&\\
&&\mathbb{S}&&\\
&\xleftarrow{\mathbb{S}}&&&\\
\hline
\end{array}
\end{displaymath}
\caption{Evaluation Phase (w.r.t. $\{V_1, \cdots V_T\}$)}
\label{fig:third-eva}
\end{scriptsize}
\end{figure}

The evaluation phase, shown in Figure \ref{fig:third-eva}, proceeds as follows.
\begin{enumerate}
\item User $i$ first establishes $M$ random messages $R_j (1 \leq j \leq M)$, random permutation functions $\mathsf{PM}$ and $\mathsf{PM}_j (1 \leq j \leq M)$, and a hash function $\mathsf{H}$ with the RecSys. Given a vector of $M$ elements, $\mathsf{PM}$ randomly permutes the order of the elements. Similarly, given a vector of $T$ elements, $\mathsf{PM}_j$ randomly permutes the order of the elements.

\item User $i$ chooses $M$ random keys $K_j (1 \leq j \leq M)$ for $\mathsf{Prf}$ and evaluates $\mathsf{Prf}$ for $R_j$ with the key $K_j - r_{j3}$ to obtain $\Upsilon_j$, for every $1 \leq j \leq M$. At the same time, the RecSys evaluates $\mathsf{Prf}$ for $R_j$ with the key $\gamma_j$ to obtain $\Theta_j$, for every $1 \leq j \leq M$.

\item After receiving the permuted values from user $i$ and the RecSys, the proxy computes
\begin{displaymath}
\Xi=\mathsf{PM}\{(\Upsilon_1, \cdots, \Upsilon_M)\} \oplus \mathsf{PM}\{(\Theta_1, \cdots, \Theta_M\}\},
\end{displaymath}
where $\oplus$ is performed element wise. It is easy to check that if the item $j$ is unrated
\begin{displaymath}
\Upsilon_j \oplus \Theta_j=\mathsf{Prf}(K_j+x_j+\epsilon_j, R_j), \mbox{and otherwise}
\end{displaymath}
\begin{displaymath}
\Upsilon_j \oplus \Theta_j=\mathsf{Prf}(K_j+r_{j1}, R_j).
\end{displaymath}

\item User $i$ first computes $\Omega_{x,j} =\mathsf{H}(\mathsf{Prf}(K_j+V_x, R_j))$ for every $1 \leq x \leq T,1 \leq j \leq M$, and then computes a randomized check value vector $\Omega_{*, j} = \mathsf{PM}_j\{(\Omega_{1,j}, \cdots, \Omega_{T,j})\}$ for every item $j$. It permutes a vector, formed by individual check value vectors of all items, and sends the result $\mathsf{PM}\{(\Omega_{*,1}, \cdots, \Omega_{*,M})\}$ to the RecSys.

\item After receiving $\mathsf{PM}\{(\Omega_{*,1}, \cdots, \Omega_{*,M})\}$ from the user, the proxy can compute $\mathbb{S}$, which is a new set generated based on $\Xi$: for every element in $\Xi$, if its hash value with respect to $\mathsf{H}$ appears in the corresponding element in $\mathsf{PM}\{\Omega_{*, j} (1 \leq j \leq M)\}$ then the corresponding element in $\mathbb{S}$ is set to be 1 otherwise it is set to be 0.

\item With $\mathbb{S}$ and $\mathsf{PM}$, user $i$ can identify the unrated items whose approximated predictions fall into the set $\{V_1, \cdots V_T\}$.
\end{enumerate}

\subsection{Security and Performance Analysis}
\label{thirdanalysis}

With encryption, the \emph{reduction} phase leaks no information to either party. In the \emph{evaluation} phase, the Recsys does not learn anything because it receives nothing from others, while user $i$ only learns which items are recommended. Regarding the information leakage to the proxy, we only need to discuss an item $j$ for any $1 \leq j \leq M$, because different $K$ and $R$ are used for different items. For any $j$, due to the fact that $K_j, r_{j3}, r_{j1}$ are chosen independently and at random, the $\Upsilon_j, \Theta_j$ are random values in the view of the proxy. With $\mathsf{H}$ being modelled as a random oracle, $\Omega_{*, j}$ leaks no information if item $j$ has been rated, and it only tells whether $\mathsf{\mathsf{Prf}(K_j+x_j+\epsilon_j, R_j)}$ is a match and nothing else. The random permutations $\mathsf{PM}$ hides which items have been recommended to the user $i$, while $\mathsf{PM}_j (1 \leq j \leq M)$ hide the predicted rating values for the recommended items. With respect to the security model from Section \ref{subsec:model}, the solution leaks the number of recommended items to the proxy, while in the security model it is required that there should be no leakage. To reduce the leakage, we can replace Step 4-6 with a privacy-preserving set interaction protocol. We leave a detailed investigation of this issue as a future work.

We summarize the asymptotic complexity in Table \ref{tab:comttp}. Based on the reference codes by the authors of \cite{CRYPTO2014a} \footnote{https://github.com/cpeikert/Lol/tree/master/lol-apps}, the $\mathsf{Prf.Evaluate}$ and $\mathsf{Prf.Hadd}$ takes about 1.04 ms and 10 $\mu$s. W.r.t. the MovieLens 1M Dataset and $T=2$, we compute the real-world running time and put it in the last column of Table \ref{tab:comttp}. It is clear that the existence of Proxy greatly improves the efficiency without seriously downgrading the privacy guarantee.


    \begin{table}[h]
        \centering
        \begin{scriptsize}
        \begin{tabular}{|c|c|c|c|c||c|}
        \hline
              &$\mathsf{Paillier.Dec}$  & $\mathsf{Paillier}.\oplus$ & $\mathsf{Prf.Evaluate}$ &$\mathsf{Prf.Hadd}$  & {\bf Time}\\
        \hline
             User &  $M$ & & $M(1 + T)$& & 63.52 s\\
        \hline
             RecSys &  & $M$ & $M$& & 4.16 s\\
        \hline
             Proxy &  & & & $M$ & 40 ms \\
        \hline
        \end{tabular}
        \end{scriptsize}
        \caption{Computational Complexities}
        \label{tab:comttp}
    \end{table}


\section{Conclusion}
\label{sec:con}

In this paper, we have demonstrated how to construct privacy-preserving collaborative filtering recommenders by separately addressing the privacy issues in the model training and prediction computation stages. We argued that the expert-based approach (e.g. \cite{ArXiv2018}) provides more scalable solution to the model training stage, while the efficiency of existing cryptographic solutions (e.g. \cite{CCS2013} and \cite{ASIACCS2016}) remains as a challenge particularly with the need to support robustness attack detection. By leveraging homomorphic encryption and key-homomorphic pseudorandom functions, we show that the proposed privacy-preserving prediction computing protocols are much more efficient than standard solutions. The current paper leaves several interesting research questions. One is to investigate the performances of cryptographic solutions when they are extended to support robustness attack detection and also improve their efficiency. Another research question is to formally study the privacy advantage of the \emph{privacy by design} approach in providing recommendations to end users, and potentially link it to differential privacy. Yet another research question is to investigate the performances (e.g. recommendation accuracy) of the two privacy-preserving protocols for the prediction computing stage based on other widely-used datasets such as Netflix.

\section*{Acknowledgement}
This work is partially funded by the European Unions Horizon 2020 SPARTA project, under
grant agreement No 830892. The author would like to thank his former colleague Jun Wang for producing Figure 3 and his current colleague Bowen Liu for running the experiment in Section 6.2.

\bibliographystyle{plain}
\bibliography{recommender}

\end{document}